\documentclass[twocolumn,english,aps,pra,reprint,floatfix,showpacs,tightenlines,superscriptaddress]{revtex4}
\usepackage[]{fontenc}
\usepackage[latin9]{inputenc}
\usepackage{amsmath}
\usepackage{graphicx}
\usepackage{amssymb}
\usepackage{esint}

\makeatletter
\@ifundefined{textcolor}{}
{%
 \definecolor{BLACK}{gray}{0}
 \definecolor{WHITE}{gray}{1}
 \definecolor{RED}{rgb}{1,0,0}
 \definecolor{GREEN}{rgb}{0,1,0}
 \definecolor{BLUE}{rgb}{0,0,1}
 \definecolor{CYAN}{cmyk}{1,0,0,0}
 \definecolor{MAGENTA}{cmyk}{0,1,0,0}
 \definecolor{YELLOW}{cmyk}{0,0,1,0}
 }

\@ifundefined{definecolor}{\@ifundefined{definecolor}{\@ifundefined{definecolor}
 {\usepackage{color}}{}
}{}}{}
\usepackage{rotating}\usepackage{enumerate}\usepackage{longtable}\usepackage{dcolumn}\usepackage{graphics}\usepackage{times}\usepackage{amsfonts}\usepackage{latexsym}\usepackage{bm}\usepackage{bbm}\usepackage{dsfont}
\usepackage{amsbsy}
\makeatletter
\@ifundefined{textcolor}{}{\definecolor{BLACK}{gray}{0}\definecolor{WHITE}{gray}{1}\definecolor{RED}{rgb}{1,0,0}\definecolor{GREEN}{rgb}{0,1,0}\definecolor{BLUE}{rgb}{0,0,1}\definecolor{CYAN}{cmyk}{1,0,0,0}\definecolor{MAGENTA}{cmyk}{0,1,0,0}\definecolor{YELLOW}{cmyk}{0,0,1,0}}

\usepackage{hyperref}

\newcommand{\tr}{\mathrm{tr}}

\newcommand{\1}{\leavevmode{\rm 1\ifmmode\mkern  -4.8mu\else\kern -.3em\fi I}}

\makeatother

\usepackage{babel}

\begin{document}

\title{Unitary equilibration after a quantum quench of a thermal state}

\author{N.~Tobias Jacobson}

\email{ntj@usc.edu}

\affiliation{Department of Physics and Astronomy and Center for Quantum Information
Science \& Technology, University of Southern California, Los Angeles,
California 90089-0484, USA}

\author{Lorenzo Campos Venuti}

\affiliation{Institute for Scientific Interchange (ISI), Viale S. Severo 65, I-10133
Torino, Italy }

\author{Paolo Zanardi}

\affiliation{Department of Physics and Astronomy and Center for Quantum Information
Science \& Technology, University of Southern California, Los Angeles,
California 90089-0484, USA}

\affiliation{Institute for Scientific Interchange (ISI), Viale S. Severo 65, I-10133
Torino, Italy }
\begin{abstract}
In this work we investigate the equilibration dynamics after a sudden
Hamiltonian quench of a quantum spin system initially prepared in
a thermal state. To characterize the equilibration we evaluate the
Loschmidt echo, a global measure for the degree of distinguishability
between the initial and time-evolved quenched states. We present general
results valid for small quenches and detailed analysis of the quantum
XY chain. The result is that quantum criticality manifests, even at
small but finite temperatures, in a universal double-peaked form of
the echo statistics and poor equilibration for sufficiently relevant
perturbations. In addition, for this model we find a tight lower bound
on the Loschmidt echo in terms of the purity of the initial state
and the more-easily-evaluated Hilbert-Schmidt inner product between
initial and time-evolved quenched states. This bound allows us to
relate the time-averaged Loschmidt echo with the purity of the time-averaged
state, a quantity that has been shown to provide an upper bound on
the variance of observables. 
\end{abstract}


\maketitle

\section{Introduction}

Consider a finite and isolated system initialized to a stationary
state of a Hamiltonian $H_{0}$. The system is then instantaneously
quenched and left to evolve unitarily according to a Hamiltonian $H_{1}$
\cite{calabrese06,Cazalilla2006,manmana07,silva08}. If we wait a
sufficiently long time, will this system ever equilibrate? It turns
out that strong equilibration cannot occur through unitary dynamics,
since no asymptotic stationary state can be reached from a non-stationary
initial state. System observables may equilibrate in a \emph{probabilistic}
sense, however, in that for the vast majority of time observable expectations
remain near to their time-averaged values. Beginning more than eighty
years ago \cite{vonneumann1929} and with resurgent interest in recent
years, a great deal of attention has been devoted to characterizing
the equilibration dynamics of such a system with the goal of deriving
from first principles the emergence of statistical mechanics \cite{tasaki98,goldstein06,reimann2,reimann08,Linden2009,Riera2011,gogolin2011,trotzky2011}.

In this paper, we investigate the effect that proximity to quantum
criticality has on the equilibration dynamics after a quench for the
realistic situation of finite temperature. To quantify the equilibration
behavior we compute the Loschmidt echo (LE), a measure for the degree
of distinguishability between initial and time-evolved quenched states
that has been used to study phenomena such as quantum chaos \cite{Peres1984,Prosen2006,Jacobson2010},
decoherence \cite{rossini_base07,rossini07,quan06,Gorin2004,Znidaric2003},
and quantum criticality \cite{quan06,PZ07,lcv2010a,lcv2010b}. In
\cite{lcv2010a,lcv2010b}, general arguments have been given, for
systems at zero temperature, that quenching near a quantum critical
point leads to poor equilibration and a universal double-peaked distribution
for observables. More recently, in Ref.~\cite{CamposVenuti2011}
we have shown that, even at finite temperature, the infinite-time
probability distribution of the Loschmidt echo takes one of two universal
forms, either double-peaked or Log-Normal. Here we will explore this
behavior in more detail. In addition, we find a tight lower bound
for the Loschmidt echo of this model in terms of a linearized quantity
that permits comparison of the time-averaged Loschmidt echo with recent
results in the statistical mechanics literature.

This paper is organized as follows: in Sec. II we introduce in general
terms the problem of equilibration for unitarily-evolving systems.
In Sec. III we discuss the Loschmidt echo for thermal states in the
general case and for small quenches in particular. In Sec.~IV we
move to consider the free fermion case and introduce the Linearized
Loschmidt Echo and related bounds. In Sec.~V we analyze in detail
the XY chain and its Ising and anisotropy transitions. Finally, Sec.~VI
contains the conclusions and the appendices detail several calculations
omitted in the main text.

\section{Unitary equilibration}

Under open quantum dynamics, it can occur that the system relaxes
towards an equilibrium state and remains near to it for (almost) all
times \cite{winter2}. However, under unitary dynamics this kind of
strong equilibration cannot happen, except for the trivial case where
the initial state is stationary. The absence of strong equilibration
can be seen clearly through the following argument. At time $t$ the
state of the system is given by $\rho\left(t\right)=U\left(t\right)\rho\left(0\right)U^{\dagger}\left(t\right)$,
with $U\left(t\right)$ the unitary quantum evolution. If $\rho\left(t\right)$
had a limit, say $\rho\left(\infty\right)$, for $t\to\infty$, this
limit must coincide with its time-average $\overline{\rho}:=\lim_{\tau\to\infty}\tau^{-1}\int_{0}^{\tau}\rho\left(t\right)dt$.
On the other hand, the invariance of $\overline{\rho}$ under the
dynamics, i.e.~$U\left(s\right)\overline{\rho}U^{\dagger}\left(s\right)=\overline{\rho}$,
implies that $\left\Vert \rho\left(t\right)-\rho\left(\infty\right)\right\Vert =\left\Vert \rho\left(t\right)-\overline{\rho}\right\Vert =\left\Vert \rho\left(0\right)-\overline{\rho}\right\Vert $
is constant for any unitary-invariant norm $\left\Vert \cdot\right\Vert $
(for example, the trace norm). But then one cannot have $\lim_{t\to\infty}\left\Vert \rho\left(t\right)-\rho\left(\infty\right)\right\Vert =0$
unless $\rho\left(t\right)=\rho\left(0\right)=\overline{\rho}$, in
which case the dynamics is trivial. This argument shows that, in the
general case $\rho\left(t\right)$ cannot have a limit in the strong
sense, i.e.~in norm. However, a weaker form of equilibration may
occur, namely the equilibration of observables in a probabilistic
sense. More precisely, we say that an observable $A$ equilibrates
in a system specified by initial condition $\rho\left(0\right)$ and
unitary evolution $U\left(t\right)$, if the expectation value $\langle A\left(t\right)\rangle:=\tr\left(\rho\left(t\right)A\right)$
stays close to its {}``equilibrium value'' for most of the times
$t$ in the observation interval $\left[0,\tau\right]$. Since the
observation time is generally much larger than the microscopic time-scales
of the dynamics, it is customary to take the limit $\tau\to\infty$.
One way to check the condition for equilibration as given above is
to consider the probability distribution $P_{A}\left(a\right):=\overline{\delta\left(\left\langle A\left(t\right)\right\rangle -a\right)}$.
$P_{A}\left(a\right)da$ gives the probability of observing a value
$\left\langle A\left(t\right)\right\rangle $ in the interval $\left[a,a+da\right]$,
during the time interval $\left[0,\tau\right]$ (and taking the limit
$\tau\to\infty$). Concentration results for $P_{A}\left(a\right)$,
i.e.~results indicating the peakedness of $P_{A}\left(a\right)$,
correspond to good equilibration. Roughly speaking, in order to show
concentration results for $P_{A}\left(a\right)$, one needs to have
access to higher and higher moments or ideally the whole distribution.
The first moment, the average value of an observable, is given in
terms of the {}``equilibrium'' state $\overline{\rho}$: $\overline{\langle A\left(t\right)\rangle}=\tr\left(\overline{\rho}A\right)$.

P. Reimann \cite{reimann08} found that the variance of an observable
can be upper-bounded in terms of the purity of the time-averaged equilibrium
state $\overline{\rho}$: \begin{equation}
\sigma_{A}^{2}\leq\Delta_{A}^{2}\mathrm{Tr}\left[\overline{\rho}^{2}\right],\label{eq:ReimannBound}\end{equation}
 where $\sigma_{A}^{2}=\overline{(\mathrm{Tr}\left[A\rho(t)\right]-\mathrm{Tr}\left[A\overline{\rho}\right])^{2}}$
is the variance of the expectation of $A$ and the \emph{range }of
$A$, $\Delta_{A}$, is a measure of the size of the observable (more
precisely, $\Delta_{A}$ is the spread of the expectations of $A$:
$\Delta_{A}=\left(a_{\mathrm{max}}-a_{\mathrm{min}}\right)$, with
respect to states in the support of the initial state $\rho(0)$ \cite{reimann08}).
As we can see, small purity of the equilibrium state implies small
variance for all observables within a given range, and hence equilibration
for a wide class of observables.

\section{The Loschmidt echo}

Given a system prepared in some initial equilibrium state $\rho$
of a Hamiltonian $H_{0}$, we would like to characterize the time-dependent
degree of distinguishability between this state and the time-evolved
states arising from the action of another Hamiltonian $H_{1}$ on
$\rho$. The quantity we consider is the Loschmidt echo, given by
\begin{equation}
\mathcal{L}(t)=F(\rho(t),\rho(0)).\label{Loschmidt_echo}\end{equation}
Here $F$ is the Uhlmann fidelity $F(\rho,\sigma)=\big(\mathrm{Tr}\sqrt{\rho^{\frac{1}{2}}\sigma\rho^{\frac{1}{2}}}\big)^{2}$,
which characterizes the degree of distinguishability between two mixed
states \cite{uhlmann76}. In this work we call $H_{0}$ ($H_{1}$)
the pre (post)-quench Hamiltonian, and take the initial state and
Hamiltonian to commute, $\left[H_{0},\rho\right]=0$.

Note that the convention we take here is to square the trace, while
in some works the Loschmidt echo is defined without the square \cite{miszczak09}.
With our convention, if $\rho$ and $\sigma$ are pure states the
Uhlmann fidelity is equivalent to a transition probability rather
than a transition amplitude. Moreover, if either (or both) of $\rho$
and $\sigma$ are pure states, then the Uhlmann fidelity simplifies
to $F(\rho,\sigma)=\mathrm{Tr}\left[\rho\sigma\right]$. In particular,
if the initial state is an eigenstate of the initial Hamiltonian,
$\rho=|\Psi_{0}\rangle\langle\Psi_{0}|$, then the Loschmidt echo
reduces to $\mathcal{L}(t)=|\langle\Psi_{0}|e^{-itH_{1}}|\Psi_{0}\rangle|^{2}$,
also known as the {}``survival probability'' in this case.

The Loschmidt echo (\ref{Loschmidt_echo}) can be equivalently expressed
in terms of a trace norm, $\mathcal{L}(t)=\Vert\rho^{\frac{1}{2}}\rho(t)^{\frac{1}{2}}\Vert_{1}^{2}$,
where $\Vert A\Vert_{1}\equiv\mathrm{Tr}|A|=\mathrm{Tr}(\sqrt{A^{\dagger}A})$.
Replacing $\Vert.\Vert_{1}$ with the Frobenius norm $\Vert.\Vert_{\mathrm{F}}$
($\Vert A\Vert_{\mathrm{F}}=\sqrt{\mathrm{Tr}\left[A^{\dagger}A\right]}$),
one obtains a simpler {}``linearized'' form of the Loschmidt echo
\begin{equation}
\mathcal{L}_{\mathrm{F}}(t):=||\rho^{\frac{1}{2}}\rho(t)^{\frac{1}{2}}||_{F}^{2}=\mathrm{Tr}(\rho(t)\rho)\leq\mathcal{L}(t),\label{general_inequality}\end{equation}
 where the bound above follows from $\Vert A\Vert_{\mathrm{F}}\le\Vert A\Vert_{1}$.
Note that this linearized quantity is generally much more convenient
to evaluate than the true Loschmidt echo, owing to the absence of
the square-roots appearing in the Uhlmann fidelity.

The aforementioned bound always holds, but we will later find a significantly
tighter lower bound on the Loschmidt echo for our model which is otherwise
valid in general only for single-qubit states (see Appendices \ref{sec:bound--qubit}
and \ref{sec:bound-XY}).

\subsection{Statistics for generic Hamiltonians\label{sec:Statistics-for-generic}}

We are interested in the equilibration properties of the Loschmidt
echo over long times. As reminded above, although the Loschmidt echo
is a purely deterministic quantity, it is useful to treat it as a
random variable in order for statistical methods to be brought to
bear. That is, we take the LE as a random variable given by its value
at a random time $t$ chosen uniformly from an interval $t\in\left[0,\tau\right]$,
and then take the limit $\tau\to\infty$. In numerical simulations
(such as the ones reported here) $\tau$ is necessarily finite. In order
to reproduce correctly the $\tau\to\infty$ limit, at least as far
as the first moments are concerned, a safe estimate is to take $\tau\gg(\min_{m,n}(E_{m}-E_{n}))^{-1}$, 
which can be as large as an exponential of the system size. This turns
out to be a far too restrictive requirement, and for the case of quasi-free
fermions analyzed here, we verified that taking $\tau\propto L^{2}$
is sufficient in order to have reliable distributions, in the sense
that larger $\tau$ produce similar distributions (see also the discussion
in Sec.~II C of Ref.~\cite{lcv2010a}). We thus consider the LE
probability distribution $P_{\mathcal{L}}\left(x\right):=\overline{\delta\left(\mathcal{L}\left(t\right)-x\right)}$,
the full-time statistics. We are interested in studying $P_{\mathcal{L}}\left(x\right)$
as the temperature and quench parameters are varied.

In the case of a pure initial state, the first moment of $P$ i.e.the
time-averaged Loschmidt echo, equals the purity of the time-averaged
state: $\overline{\mathcal{L}}=\mathrm{Tr}\left[\overline{\rho}^{2}\right]$.
Hence from Eq (\ref{eq:ReimannBound}) it follows that for this case
a smaller $\overline{\mathcal{L}}$ corresponds to a smaller variance
(\ref{eq:ReimannBound}) for generic observables and hence {}``good''
quantum equilibration. However, in the case we are considering here
where the initial state can be mixed, we will see that the LE need
not be small for the equilibrium purity to be small.

In Ref.~\cite{CamposVenuti2011}, we found that the LE takes one
of two universal forms, either Log-Normal in the off-critical region
or double-peaked in the quasi-critical region. That is, considering
the variable $\mathcal{Z}\left(t\right):=\ln\mathcal{L}\left(t\right)$,
the distribution for the \emph{logarithm} of the LE, $P_{\mathcal{Z}}\left(x\right)$
is then correspondingly either Gaussian or double-peaked. We obtained
this result for a quasi-free system through a central limit theorem
(CLT)-type argument for $\mathcal{Z}$. In Ref.~\cite{CamposVenuti2011},
we also argued that such universal statistics should hold as well for
small quenches with generic (i.e.~not necessarily quasi-free) Hamiltonians
in the pure case, i.e.~at zero temperature.

\subsection{Small quenches}

We now provide an argument in favor of the preceding result also in
the mixed, i.e.~non-zero temperature, case. Namely, for small quenches
of generic Hamiltonians and thermal initial states proximity to quantum
criticality gives rise to double-peaked statistics, whereas otherwise,
in the off-critical region, one expects Gaussian behavior for $P_{\mathcal{Z}}(x)$.

Suppose our post-quench Hamiltonian is $H_{1}=\sum_{n}E_{n}^{1}\vert\psi_{n}\rangle\langle\psi_{n}\vert$
and our initial state is given by $\rho(0)=\sum_{n,m}\rho_{n,m}\vert\psi_{n}\rangle\langle\psi_{m}\vert$.
We assume that this state is diagonal in the pre-quench Hamiltonian's
eigenbasis, $\rho(0)=\sum_{n}p_{n}\vert n\rangle\langle n\vert$,
where we take $\lbrace\vert n\rangle\rbrace$ ($\lbrace\vert\psi_{n}\rangle\rbrace$)
to correspond to the pre (post)-quench Hamiltonian eigenstates. For
nearby states $\rho,\rho+\delta\rho$, the lowest-order expansion
of the Uhlmann fidelity is given by \cite{Hubner1992,Sommers2003}
\begin{equation}
F(\rho,\rho+\delta\rho)=1-\frac{1}{2}\sum_{m,n}\frac{\vert\langle m\vert\delta\rho\vert n\rangle\vert^{2}}{p_{m}+p_{n}}.\label{eq:BuresFid}\end{equation}

In our case $\delta\rho$ will be time-dependent and correspond to
the difference between $\rho(t)$ and $\rho(0)$. In order to use
the expansion Eq.~(\ref{eq:BuresFid}), we have to show that $\delta\rho\left(t\right)$
is small in $V$ \emph{independently} of $t$, since we will use such
an expansion for any time $t$. To see this, compute $\delta\rho\left(t\right)$
exactly in the post-quench eigenbasis: \begin{eqnarray}
\delta\rho\left(t\right) & = & \rho(t)-\rho(0)\label{eq:DeltaRho}\\
 & = & \sum_{n\neq m}\rho_{n,m}\left(e^{-it\left(E_{n}^{1}-E_{m}^{1}\right)}-1\right)\vert\psi_{n}\rangle\langle\psi_{m}\vert.\nonumber \end{eqnarray}
 Observe that $\delta\rho\left(t\right)$ has only off-diagonal elements
in the post-quench basis. Now using time-independent perturbation
theory (and assuming a non-degenerate spectrum for simplicity), one
can show that for $n\neq m$, $\left|\rho_{n,m}\right|\le\mathrm{const.}\times V$
(where for simplicity of notation $V$ denotes also the strength of
the perturbation). By then bounding the time-oscillating terms by
two, we obtain $\left\Vert \delta\rho\left(t\right)\right\Vert \le O\left(V\right)$
in some norm, independent of $t$. This argument indicates that one
can use a Dyson expansion for any arbitrary time $t$.

Evaluating the matrix element $\langle m\vert\delta\rho\left(t\right)\vert n\rangle$
up to first order in the perturbation using Eqs.~(\ref{eq:BuresFid})
and (\ref{eq:DeltaRho}) we obtain, to second order in the perturbation

\begin{eqnarray}
\mathcal{L}(t) & \simeq & 1-\sum_{n\neq m}C_{n,m}\left[1-\cos\left[(E_{m}^{1}-E_{n}^{1})t\right]\right],\label{eq:GeneralLog}\end{eqnarray}
 where \begin{equation}
C_{n,m}=\frac{(p_{m}-p_{n})^{2}}{p_{m}+p_{n}}\frac{\vert V_{m,n}\vert^{2}}{(E_{m}^{0}-E_{n}^{0})^{2}}\end{equation}
 and $V_{m,n}=\langle m\vert V\vert n\rangle$. We can now take the
time-average of Eq.~(\ref{eq:GeneralLog}) with any arbitrarily-large
observation time $\tau$ (i.e.~also $\tau\to\infty$), with the result
that the time-averaged Loschmidt echo is given by \begin{equation}
\overline{\mathcal{L}}=1-\sum_{n\neq m}\frac{(p_{m}-p_{n})^{2}}{p_{m}+p_{n}}\frac{\vert V_{m,n}\vert^{2}}{(E_{m}^{0}-E_{n}^{0})^{2}}.\label{eq:ZBarSum}\end{equation}

We next show how this time-average is related with the state fidelity.
Let us then consider, in general, the Uhlmann fidelity $F(\rho_{0},\rho_{1})$
between two nearby states $\rho_{0}$, $\rho_{1}=\rho_{0}+d\rho$.
For example, $\rho_{0}$ ($\rho_{1}$) can be a thermal Gibbs state
relative to the pre (post)-quench Hamiltonian $H_{0}$ ($H_{1}$).
For small quenches $d\rho=O\left(V\right)$ and we can use the general
expansion Eq.~(\ref{eq:BuresFid}) which defines the Bures distance
$ds^{2}$. The variation $d\rho$ in this case takes contributions
both from the change of eigenvalues and eigenvectors, $d\rho=\sum_{n}(d\rho_{n}\vert n\rangle\langle n\vert+p_{n}(\vert dn\rangle\langle n\vert+\vert n\rangle\langle dn\vert)$.
Differentiating the eigenvalue equation $H\vert n\rangle=E_{n}\vert n\rangle$
in order to evaluate $\langle m\vert dn\rangle$, we obtain \cite{ZanardiCamposVenuti2007}
\begin{align}
F & =1-ds^{2}\nonumber \\
ds^{2} & =\frac{1}{4}ds_{\mathrm{FR}}^{2}+\frac{1}{2}\sum_{n\neq m}\frac{(p_{m}-p_{n})^{2}}{p_{m}+p_{n}}\frac{\vert\langle m\vert dH\vert n\rangle\vert^{2}}{(E_{n}-E_{m})^{2}}.\label{eq:FinalMetric}\end{align}
 Here $ds^{2}$ is the Bures metric, while $ds_{\mathrm{FR}}^{2}$
is the so-called Fisher-Rao distance between probability distributions
corresponding to the sets of diagonal weights $\lbrace p_{n}\rbrace$
and $\lbrace p_{n}+dp_{n}\rbrace$ and is given by $ds_{\mathrm{FR}}^{2}=\sum_{n}dp_{n}^{2}/p_{n}$.
The second term in Eq.~(\ref{eq:FinalMetric}) is what is denoted
the {}``non-classical'' part and depends on the variation of the
eigenstates \cite{ZanardiCamposVenuti2007}. Through comparing Eqns.~(\ref{eq:FinalMetric})
and (\ref{eq:ZBarSum}), we can establish the relation $\left[F\left(\rho_{0},\rho_{1}\right)\right]^{2}=\overline{\mathcal{L}}-ds_{\mathrm{FR}}^{2}/2$,
valid up to second order in the perturbation. This result provides
a relation between the Loschmidt echo (a dynamical quantity) and the
state fidelity (a static one). For pure initial states (when the {}``temperature''
goes to zero) one recovers $F^{2}=\overline{\mathcal{L}}$ \cite{rossini_base07},
since in this limit the classical part of the Bures metric which depends
on the variation of the eigenvalues vanishes \cite{ZanardiCamposVenuti2007}.

Let us now analyze the full time statistics and consider the logarithm
of the Loschmidt echo for small quenches, Eq.~(\ref{eq:GeneralLog}).
Assuming the initial state is of the Gibbs form $\rho(0)=e^{-\beta H_{0}}/Z$,
the weights $p_{n}$ are given by $p_{n}=e^{-\beta E_{n}^{0}}/Z$.
This means that the largest weights $C_{n,m}$ are to be found when
either $n$ or $m$ is zero, since $C_{n,m}\ll C_{n,0}$ for all $m\neq0$,
a result which is (exponentially) stronger the lower the temperature.
For temperatures sufficiently low as compared to the gaps, i.e.~$\beta(E_{1}^{0}-E_{0}^{0})\gg1$,
a good approximation to $\ln\mathcal{L}\left(t\right)$ is then given
by (taking $W_{n,m}=2C_{n,m}$) \begin{equation}
\mathcal{Z}\left(t\right)\simeq\overline{\mathcal{Z}}+\sum_{n>0}W_{n,0}\cos((E_{n}^{0}-E_{0}^{0})t),\label{eq:Zt}\end{equation}
with $\overline{\mathcal{Z}}=-\sum_{n\neq m}C_{n,m}$. 

If we now assume that the perturbed energy gaps $\Delta_{n}:=(E_{n}^{0}-E_{0}^{0})$
are \emph{rationally independent} (i.e.~linearly independent over
the field of rational numbers), using the ergodic theorem one realizes
that $\mathcal{Z}\left(t\right)-\overline{\mathcal{Z}}$ is a \emph{sum
}of \emph{independent }random variables each distributed according
to $P\left(z_{n}\right)=\vartheta\left(W_{n,0}^{2}-z_{n}^{2}\right)/\left(\pi\sqrt{W_{n,0}^{2}-z_{n}^{2}}\right)$
\cite{CamposVenuti2011}. To see this let us compute the characteristic
function of the variable $\mathcal{Z}-\overline{\mathcal{Z}}$, $\overline{e^{i\lambda\left(\mathcal{Z}-\overline{\mathcal{Z}}\right)}}$.
Rational independence allows to invoke the ergodic theorem and to
compute the time expectation value as an average over a multidimensional
torus with uniform measure:\begin{equation}
\overline{e^{i\lambda\left(\mathcal{Z}-\overline{\mathcal{Z}}\right)}}=\prod_{n>0}\int_{0}^{2\pi}\frac{d\vartheta_{n}}{2\pi}e^{i\lambda W_{n,0}\cos\left(\vartheta_{n}\right)}.\label{eq:char_general}\end{equation}
 This also shows that $\mathcal{Z}$ is a sum of independent variables
since the characteristic function factorizes into a product of functions.
The integration over each angle gives a Bessel function $J_{0}\left(\left|\lambda W_{n,0}\right|\right)$.
The probability distribution of each mode is computed by Fourier transforming
this Bessel function and gives rise to the distribution $P\left(z_{n}\right)$
given above. Each $P\left(z_{n}\right)$ is the density of states
(DOS) of a one-dimensional tight-binding model with coupling $W_{n,0}/2$
and has mean zero and variance given by $\left(W_{n,0}\right)^{2}/2$.
The collective distribution of $\mathcal{Z}\left(t\right)-\overline{\mathcal{Z}}$
is also given, in principle, by the DOS of a huge-dimensional ($D-1$,
where $D$ is the Hilbert space dimension) tight-binding model with
anisotropic couplings $W_{n,0}/2$ in each direction. Given that the
size of $D$ is large, the distribution of $\mathcal{Z}$ is better
computed using a central limit theorem-type argument. Therefore we
need to control the weights $W_{n,0}$ which determine the variance
of $\mathcal{Z}$.

\subsection{Temperature and criticality\label{sub:Temperature-and-criticality}}

The weights $W_{n,0}$ are simply temperature-damped versions of their
corresponding $T=0$ values: $W_{n,0}\left(T\right)=\mathcal{D}_{n}\left(T\right)W_{n,0}\left(T=0\right)$,
where \[
\mathcal{D}_{n}\left(T\right)=\frac{e^{-\beta E_{0}^{0}}}{Z}\frac{\left(e^{-\beta\Delta_{n}}-1\right)^{2}}{e^{-\beta\Delta_{n}}+1}\]
 and $W_{n,0}\left(0\right)=2\vert V_{n,0}\vert^{2}/(E_{n}^{0}-E_{0}^{0})^{2}$.
The temperature damping factor $0\le\mathcal{D}_{n}\left(T\right)\le1$
has simply the effect of attenuating the level $n$ as temperature
increases.

Let us first review the situation at $T=0$. The factor $W_{n,0}\left(0\right)$
has already been considered in \cite{lcv2010b,CamposVenuti2011,polkovnikov09,polkovnikov10}.
It was found that close to quantum criticality, $W_{n,0}$ scales
with the energy as $E^{-2/\left(\zeta\nu\right)}$ , where $\zeta$
is the dynamical critical exponent and $\nu$ is the correlation length
one. The algebraic divergence at low energy of $W_{n,0}\left(0\right)$
has the effect that, sufficiently close to criticality (i.e.~when
the correlation length $\xi$ is much larger than the linear system
size $L$), few low-energy weights $W_{n,0}\left(0\right)$ absorb
most of the total weight \cite{lcv2010b}. In turn this implies that,
at $T=0$, the sum over $n$ in Eq.~(\ref{eq:Zt}) can be safely
restricted to a small number $n_{\mathrm{max}}$ of energy levels.
As a result, the distribution of $\mathcal{Z}$ is the DOS of an $n_{\mathrm{max}}$-dimensional
tight-binding model. For small $n_{\mathrm{max}}$ (i.e.~$n_{\mathrm{max}}\le2\div3$)
such a distribution is a double-peaked function with a large variance
\cite{lcv2010a,lcv2010b} corresponding to a universal kind of poor
equilibration.

We can now ask ourselves how this result will be modified when switching
on the temperature $T>0$. If $\Delta$ ($\xi$) denotes the energy
gap above the ground state (correlation length) one has to consider
different cases 
\begin{itemize}
\item [a)] Region I: $\xi\gg L<\infty$ and $\Delta\ll T=\beta^{-1}$
(in any case the temperature must still be sufficiently small with
respect to the energy scale of the problem). The condition $\Delta/T\ll1$
allows to expand the temperature-damping factor $\mathcal{D}_{n}\left(T\right)$
as $\mathcal{D}_{n}\left(T\right)\simeq E^{2}/T^{2}$. Combining this
result with the zero-temperature scaling we obtain \[
W_{n}\left(T\right)\simeq E^{2-2/\left(\zeta\nu\right)}/T^{2}\]
 The divergence observed at zero temperature is softened by the factor
$E^{2}$, but there still may be situations where $W_{n}\left(T\right)$
diverges at low energy, namely for $\zeta\nu<1$. In other words,
in the quasi-critical region the possibility of observing a double-peaked
distribution for $\mathcal{Z}$ depends on the relevance of the perturbation.
For sufficiently relevant perturbation, $\zeta\nu<1$ (or differently
put $\Delta_{V}<d$, where $d$ is the spatial dimension and $\Delta_{V}$
the scaling dimension of $V$) the distribution of $\mathcal{Z}$
is double-peaked, whereas for $\zeta\nu\ge1$ the expected behavior
is the Gaussian one corresponding to {}``good'' equilibration.
\item [b)] Region II: $\xi\gg L$ but $\Delta\gg T$. The condition $\Delta/T\gg1$
fixes $\mathcal{D}_{n}\left(T\right)$ to one up to exponentially
small corrections, and we recover the $T=0$ case, namely a double-peaked
distribution for $\mathcal{Z}$ corresponding to poor equilibration.
\item [c)] Region III (off-critical region, thermodynamic limit): $L\gg\xi$.
The variance of $\mathcal{Z}$ is given by $\kappa_{2}\left(\mathcal{Z}\right)=\left(1/2\right)\sum_{n}\left[W_{n,0}\left(T\right)\right]^{2}$.
For a sufficiently small perturbation both $\mathcal{D}_{n}\left(T\right)$
and $W_{n,0}\left(0\right)$ will be between zero and one. Then \[
2\kappa_{2}\left(\mathcal{Z}\right)\le\sum_{n}\left[W_{n,0}\left(0\right)\right]^{2}\le\sum_{n}W_{n,0}\left(0\right)\equiv\chi_{F}\]
 The quantity $\chi_{F}$ is the (zero-temperature) {}``fidelity
susceptibility'' and was shown to grow at most extensively in the
off -critical region \cite{lcv07}. Hence $\kappa_{2}\left(\mathcal{Z}\right)\le O\left(L\right)$
and the rescaled variable $\mathcal{Y}:=\left(\mathcal{Z}-\overline{\mathcal{Z}}\right)/\sqrt{L}$
tends in distribution, as $L\to\infty$, to a Gaussian with zero mean
and variance $\sigma^{2}=\lim_{L\to\infty}\kappa_{2}\left(\mathcal{Z}\right)/L$
(the limit exists because $\kappa_{2}\left(\mathcal{Z}\right)$ is
monotonically increasing with $L$, since $\mathcal{Z}$ is a sum
of independent variables).
\item [d)] Finally, for the sake of completeness, we analyze the high-temperature
region. For $\Delta_{n}/T\ll1\,(\forall n)$ one has $\mathcal{D}_{n}\left(T\right)=O(T^{-2})$
and therefore $\mathcal{L}(t)=1-O(T^{-2}).$ This is of course just
a consequence of the fact that the thermal state approaches the maximally
mixed one for infinite temperature and the latter has trivial dynamics
(for all Hamiltonians).
\end{itemize}

\section{Quasi-free fermions}

We now turn our attention to a special class of systems for which
a closed-form expression for the Loschmidt echo has been found \cite{PZ07}:
those systems described by a quasi-free fermion Hamiltonian \begin{equation}
H=\sum_{k}\epsilon_{k}(c_{k}^{\dagger}c_{k}+c_{-k}^{\dagger}c_{-k})+\Delta_{k}(-ic_{k}^{\dagger}c_{-k}^{\dagger}+ic_{-k}c_{k}),\label{eq:Quasi-Free}\end{equation}
 where $c_{k}^{\dagger}$ ($c_{k}$) create (annihilate) spinless
fermions. Such a model can be recast in a diagonal form $H=\sum_{k}\Lambda_{k}\eta_{k}^{\dagger}\eta_{k}$,
where $\lbrace\Lambda_{k}\rbrace$ are single-particle energies. Imposing
anti-periodic boundary conditions, the quasi-momenta $k$ are quantized
according to $k=(2n+1)\pi/L$, where $n=0,\dots,L-1$ and we will
assume $L$ to be even.

The Gibbs (thermal) state of such a model at inverse temperature $\beta$
takes the form of a tensor-product of $4\times4$ density matrices
\cite{PZ07}, \begin{equation}
\rho=\frac{e^{-\beta H}}{Z}=\bigotimes_{k>0}\frac{1}{Z_{k}}\Big(\rho_{k}^{\textrm{even}}\oplus\1_{k}^{\textrm{odd}}\Big),\label{Gibbs_state}\end{equation}
 where $\rho_{k}^{\textrm{even}}$ is a $2\times2$ matrix over the
{}``even'' subspace spanned by $\lbrace\vert0\rangle,c_{k}^{\dagger}c_{-k}^{\dagger}\vert0\rangle\rbrace$
and $\1_{k}^{\textrm{odd}}$ is the $2\times2$ identity operator
over the {}``odd'' subspace $\lbrace c_{k}^{\dagger}\vert0\rangle,c_{-k}^{\dagger}\vert0\rangle\rbrace$.
The tensor product is over $L/2$ momentum modes. This splitting of
a given momentum mode into an even and odd subspace is a consequence
of the Hamiltonian acting only trivially (as an energy shift) on the
odd subspace \cite{PZ07}.

For any quench Hamiltonian $H$ of the form (\ref{eq:Quasi-Free})
the unitary operator $U(t)=e^{-itH}$ is directly analogous to (\ref{Gibbs_state}),
as can be seen readily by performing a Wick rotation $\beta\to it$.

Exploiting the special factorized form of the Gibbs state and the
evolution operator, we are able to prove the following bound (see
Appendix \ref{sec:bound-XY}). \begin{equation}
\frac{\mathrm{Tr}(U\rho U^{\dagger}\rho)}{\mathrm{Tr}(\rho^{2})}\leq\Big(\mathrm{Tr}\sqrt{\rho^{\frac{1}{2}}U\rho U^{\dagger}\rho^{\frac{1}{2}}}\Big)^{2}.\label{eq:qubit_inequality}\end{equation}
 This inequality provides a much-tighter lower bound on the Loschmidt
echo than the general inequality (\ref{general_inequality}). Note
that $U\to\1$ gives unity on both sides of the inequality. The inverse
purity is usually given the name of \emph{effective dimension}, $d_{\mathrm{eff}}:=1/\mathrm{Tr(\rho^{2})}$
\cite{Linden2009}. The effective dimension provides a weighted measure
of how broadly the ensemble weights of $\rho$ are distributed \cite{Linden2009}.
In particular, the effective dimension of a pure state is unity while
that of a maximally-mixed state is equal to the Hilbert space dimension.
Inequality (\ref{eq:qubit_inequality}) may then be expressed equivalently
as \begin{equation}
d_{\mathrm{eff}}\mathcal{L}_{\mathrm{F}}(t)\leq\mathcal{L}(t).\label{eq:Lower_bound}\end{equation}
 The tightness of this bound with respect to (\ref{general_inequality})
can be appreciated since $1\leq d_{\mathrm{eff}}\leq d$, where $d$
is the Hilbert space dimension. Again, we stress that (\ref{eq:Lower_bound})
is not generally applicable to states belonging to a Hilbert space
of dimension greater than two. We remark that the quantity on the
left-hand side of inequality (\ref{eq:Lower_bound}) was proposed
by Peres in Ref.~\cite{Peres1984} as a generalization of the Loschmidt
echo to mixed states.

\begin{figure}[hbp]
 \centering{} \includegraphics[scale=0.35]{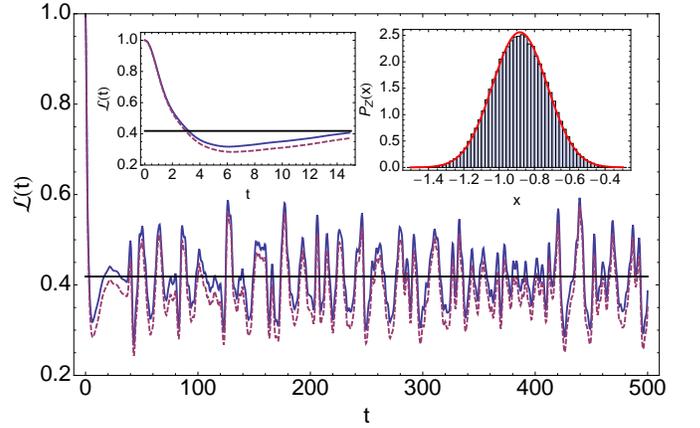} \caption{\label{fig:IsingEcho_L20} (Color online) Time series of $\mathcal{L}(t)$
(solid) and $d_{\mathrm{eff}}\mathcal{L}_{\mathrm{F}}(t)$ (dashed)
along with the average $\overline{\mathcal{L}}$ for the quantum XY
chain, with $L=80$, $h^{0,1}=0.5$, $\gamma^{0}=0.25$, $\gamma^{1}=0.1$
and $\beta=10$. In the left inset the initial Gaussian decay is evident.
The right inset shows a histogram of $\mathcal{Z}=\ln{\mathcal{L}}$
with a superimposed Gaussian distribution (red) of the same mean and variance.}

\end{figure}

To bound $\mathcal{L}(t)$ from above we employ the following generally-applicable
inequality in terms of the so-called \emph{super-fidelity} due to
Miszczak, et al.~\cite{miszczak09} \begin{equation}
\mathrm{Tr}\left[\sqrt{\rho^{\frac{1}{2}}\sigma\rho^{\frac{1}{2}}}\right]^{2}\leq\mathrm{Tr}\left[\rho\sigma\right]+\sqrt{(1-\mathrm{Tr}\rho^{2})(1-\mathrm{Tr}\sigma^{2})}.\end{equation}
 Putting this all together, we can now bound the Loschmidt echo from
below and above in terms of the linearized echo: \begin{equation}
d_{\mathrm{eff}}\mathcal{L}_{\mathrm{F}}(t)\leq\mathcal{L}(t)\leq\mathcal{L}_{\mathrm{F}}(t)+\Big(1-d_{\mathrm{eff}}^{-1}\Big).\label{tight_inequality}\end{equation}
 Observe that these bounds become tighter as the purity of the initial
state gets closer to unity. This corresponds, in our setting, to the
low-temperature regime. Note that for $t=0$ the upper and lower bounds
are equal, so by continuity they are expected to characterize well
the short-time behavior of $\mathcal{L}(t)$. Typical behavior of
$\mathcal{L}(t)$ and $d_{\mathrm{eff}}\mathcal{L}_{\mathrm{F}}(t)$
is depicted in Fig. \ref{fig:IsingEcho_L20}. The Loschmidt echo drops
from unity at $t=0$ and then oscillates about its average value,
with almost periodic revivals \cite{alioscia_revival}. We find that,
at each fixed time, $\mathcal{L}$ monotonically increases with temperature,
which can be understood as being due to the Gibbs state tending towards
the totally-mixed state $\1/d$ for increasing temperature.

\subsection{Loschmidt Echo Statistics}

For model (\ref{eq:Quasi-Free}), the Loschmidt echo has been shown
to be \cite{PZ07} \begin{equation}
\mathcal{L}(t)=\prod_{k>0}\left[\frac{1+\sqrt{c_{k}^{2}-(c_{k}^{2}-1)\alpha_{k}\sin^{2}(\Lambda_{k}^{1}t)}}{1+c_{k}}\right]^{2},\label{eq:XY_Loschmidt_echo}\end{equation}
 where $c_{k}=\cosh(\beta\Lambda_{k}^{0})$, $\alpha_{k}=\sin^{2}(\Delta\theta_{k})$,
$\Delta\theta_{k}=\theta_{k}^{1}-\theta_{k}^{0}$, and $\theta_{k}=\arctan(\Delta_{k}/\epsilon_{k})$.

The linearized LE and the effective dimension have similar, though
simpler, product forms (see Appendices \ref{sec:Linearized-Loschmidt-echo}
and \ref{sec:Effective-dimension}): \begin{equation}
\mathcal{L}_{\mathrm{F}}(t)=d_{\mathrm{eff}}^{-1}\prod_{k>0}\left[1-(1-c_{k}^{-2})\alpha_{k}\sin^{2}(\Lambda_{k}^{1}t)\right]\end{equation}
 and \begin{equation}
d_{\mathrm{eff}}^{-1}=\mathrm{Tr}\left[\rho^{2}\right]=\prod_{k>0}\left[c_{k}/(c_{k}+1)\right]^{2}.\label{Gibbs_purity}\end{equation}
 Since $c_{k}$ increases monotonically with the temperature, the
effective dimension ranges continuously and monotonically, from $d_{\mathrm{eff}}=1$
at $T=0$ to $d_{\mathrm{eff}}=d$, the Hilbert space dimension, when
$T\to\infty$.

The time average of (\ref{eq:XY_Loschmidt_echo}) is (see Appendix
\ref{sec:Computation-of-averages}): \begin{equation}
\overline{\mathcal{L}}=\prod_{k>0}\left[1-(1-c_{k}^{-1})\frac{\alpha_{k}}{2}+g_{k}\right],\end{equation}
 where $g_{k}\equiv\frac{2c_{k}}{(1+c_{k})^{2}}\left[\frac{2}{\pi}\mathrm{E}(-b_{k})-\frac{b_{k}}{4}-1\right]$
and $\mathrm{E}(x)$ is the complete elliptic integral.

The time-averaged linearized echo, $\overline{\mathcal{L}_{\mathrm{F}}(t)}=\mathrm{Tr}\left[\overline{\rho(t)}\rho\right]$,
is easier to evaluate. Assuming rational independence of half of the
single-particle energies, $\left\{ \Lambda_{k}\right\} _{k>0}$ %
\footnote{In most cases this seems to be a plausible assumption, see note {[}13{]}
of Ref.~\cite{CamposVenuti2011}.%
}, the time-average is \begin{equation}
\overline{\mathcal{L}_{\mathrm{F}}(t)}=d_{\mathrm{eff}}^{-1}\prod_{k>0}\left[1-(1-c_{k}^{-2})\frac{\alpha_{k}}{2}\right].\label{eq:AvgLEF}\end{equation}
 As we can see in Eq.~(\ref{eq:AvgLEF}), to obtain a smaller $\overline{\mathcal{L}_{\mathrm{F}}}$
we can either increase the effective dimension (by increasing the
temperature, for example) or enhance the strength of the quench.

Note that the equilibrium state $\overline{\rho}$ is diagonal in
the eigenbasis of the quench Hamiltonian. To see this, expand the
initial state in terms of the quench Hamiltonian eigenbasis and take
the time average, exploiting the non-degeneracy of the spectrum (see
e.g.~\cite{Linden2009}). The equilibrium state is then only the
diagonal part of the initial state in the quench basis. Consequently,
\begin{equation}
\overline{\mathcal{L}_{\mathrm{F}}(t)}=\mathrm{Tr}\left[\overline{\rho}^{2}\right],\end{equation}
 i.e.~the time-averaged linearized echo is simply the purity of the
equilibrium state. Plugging this into the inequality (\ref{tight_inequality}),
we obtain \begin{equation}
\mathrm{Tr}\left[\overline{\rho}^{2}\right]\mathrm{Tr}\left[\rho^{2}\right]^{-1}\leq\overline{\mathcal{L}(t)}\leq1-\big(\mathrm{Tr}\left[\rho^{2}\right]-\mathrm{Tr}\left[\overline{\rho}^{2}\right]\big).\end{equation}
 This result says that the time-averaged Loschmidt echo is bounded
from below by the ratio of the effective dimensions of the initial
and equilibrium states and the difference from unity from above is
given by at least the difference between initial and equilibrium purities.
Note that in the limit of a \emph{pure} (unity purity) initial state,
both lower and upper bounds would be equal and we would recover the
well-known result \cite{Campos2010} $\overline{\mathcal{L}}=\mathrm{Tr}\left[\overline{\rho}^{2}\right]$.

At this point, we can make a connection between the time-averaged
Loschmidt echo and the bound on observable variances found by Reimann,
Eq.~(\ref{eq:ReimannBound}). Since the upper bound is in terms of
the purity of the time-averaged state (i.e. the time-averaged linearized
echo), it is clear that the average Loschmidt echo need not be small
to have good equilibration. Indeed, provided the initial purity $\mathrm{Tr}\left[\rho(0)^{2}\right]$
is small enough, $\overline{\mathcal{L}}$ can be as close as we wish
to unity. On the other hand, if $\overline{\mathcal{L}}$ is small,
we are guaranteed that $\mathrm{Tr}\left[\overline{\rho}^{2}\right]$
will be even smaller, since $\mathrm{Tr}\left[\overline{\rho}^{2}\right]\leq\overline{\mathcal{L}}/d_{\mathrm{eff}}$.

To find the short-time behavior of the Loschmidt echo, it is convenient
to express the linearized echo as \begin{equation}
\mathcal{L}_{\mathrm{F}}(t)=\ll\!\rho|e^{-it\mathcal{H}}|\rho\!\gg,\label{superoperator_echo_rep}\end{equation}
 where $\mathcal{H}$ is a superoperator acting on elements of the
Hilbert-Schmidt space as $\mathcal{H}\alpha=\left[H,\alpha\right]$.
$|\rho\!\!\gg$ is a ket in the Hilbert-Schmidt space, equivalent
to the density operator $\rho$. Note, however, that this ket is not
normalized with respect to the Hilbert-Schmidt norm, i.e.~$\ll\!\!\rho|\rho\!\!\gg=\mathrm{Tr}\left[\rho^{2}\right]=d_{\mathrm{eff}}^{-1}\neq1$
in general. Eq.~(\ref{superoperator_echo_rep}) is the direct analog
of the expression for the Loschmidt echo at zero temperature, $\mathcal{L}(t)=|\langle\Psi|e^{-itH}|\Psi\rangle|^{2}$.

The linearized echo can alternatively be expressed as the Fourier
transform of the energy gap probability distribution $\chi(\omega):=\langle\delta(\omega-\mathcal{H})\rangle=\ll\!\!\rho|\delta(\omega-\mathcal{H})|\rho\!\!\gg$.
Using this, in the same spirit as the analogous zero-temperature analysis
\cite{lcv2010a} we consider the cumulant expansion of $\mathcal{L}_{\mathrm{F}}(t)$
\begin{equation}
\mathcal{L}_{\mathrm{F}}(t)=d_{\mathrm{eff}}^{-1}\exp\left[\sum_{n=2}^{\infty}\frac{(-it)^{n}}{n!}\ll\!\!\mathcal{H}^{n}\!\!\gg_{c}\right],\end{equation}
 where $\ll\!\!\mathcal{H}^{n}\!\!\gg_{c}$ represents the $n$-th
cumulant of $\mathcal{H}$ with respect to the state $\rho$ and $d_{\mathrm{eff}}^{-1}$
is the normalization factor to ensure $\mathcal{L}_{\mathrm{F}}(0)=\mathrm{Tr}\rho^{2}$.
The sum begins at $n=2$ due to the cyclicity of the trace. The second
cumulant is the variance of $\mathcal{H}$, \begin{eqnarray*}
\ll\!\!\mathcal{H}^{2}\!\!\gg_{c} & = & 2\big(\mathrm{Tr}\left[\rho^{2}H^{2}\right]-\mathrm{Tr}\left[(\rho H)^{2}\right]\big)\\
 & = & 2\mathrm{Tr}\left[\rho^{2}\right]\sum_{k>0}\frac{\mathrm{Tr}\left[\rho_{k}^{2}H_{k}^{2}\right]-\mathrm{Tr}\left[(\rho_{k}H_{k})^{2}\right]}{\mathrm{Tr}\left[\rho_{k}^{2}\right]+2},\end{eqnarray*}
 and for short times, where only the second-order term contributes,
we have \begin{equation}
d_{\mathrm{eff}}\mathcal{L}_{\mathrm{F}}(t)\approx1-\frac{1}{2}\ll\!\!\mathcal{H}^{2}\!\!\gg_{c}t^{2}.\end{equation}
 Checking both this expansion and the short-time expansion of the
exact Loschmidt echo (\ref{eq:XY_Loschmidt_echo}), we find that in
the off-critical region $1-\mathcal{L}(t)\propto t^{2}L$, a scaling
that coincides with the zero-temperature result \cite{lcv2010a}.

\section{The Quantum XY chain}

An important instance of the above class of quasi-free fermions (\ref{eq:Quasi-Free})
is the quantum XY chain in a transverse magnetic field, \begin{equation}
H=-\sum_{i=1}^{L}\frac{(1+\gamma)}{2}\sigma_{i}^{x}\sigma_{i+1}^{x}+\frac{(1-\gamma)}{2}\sigma_{i}^{y}\sigma_{i+1}^{y}+h\sigma_{i}^{z}.\label{eq:XY_Hamiltonian}\end{equation}
 This is a well-known and long-studied model. A Jordan-Wigner mapping
and Fourier transform bring the model to the form of Eq.~(\ref{eq:Quasi-Free}), 
which allows for an exact solution \cite{LiebSchultzMattis1961,Pfeuty1970}.
The XY chain (\ref{eq:XY_Hamiltonian}) is of the same form as (\ref{eq:Quasi-Free})
with the identification $\epsilon_{k}=\cos(k)+h$ and $\Delta_{k}=\gamma\sin(k)$.
The single-particle energies and the eigenstate-parametrizing angles
are defined as previously: $\Lambda_{k}=\sqrt{\epsilon_{k}^{2}+\Delta_{k}^{2}}$
and $\tan(\theta_{k})=\frac{\Delta_{k}}{\epsilon_{k}}$. As is customary, 
we fix boundary conditions on the Fermi operators and choose anti-periodic
ones \cite{lsm61,barouch70}. The momenta then take the values $k=(2n+1)\pi/L$,
with $n=0,\dots,L-1$.

This system exhibits two types of quantum critical lines at zero temperature:
(i) the Ising transition for $\vert h\vert=1$ and $\gamma\neq0$,
corresponding to a second-order quantum phase transition between ferromagnetic
and paramagnetic phases, and (ii) the anisotropy transition for $\gamma=0$
and $\vert h\vert<1$, a second-order QPT between ferromagnetic phases
with long-range order in the $x$ or $y$ directions for $\gamma>0$
or $\gamma<0$, respectively.

Here, we will explore in more detail the conditions for obtaining
either of the two universal distributions for the LE (Log-Normal or
double-peaked) for small quenches of the quantum XY chain.

In the following, we consider the logarithm of the LE, denoted by
the quantity $\mathcal{Z}(t)=\ln\mathcal{L}(t)$. Expanding $\mathcal{Z}(t)$
up to second order in the quench amplitude, we obtain 
\begin{equation}
\mathcal{Z}(t)=\overline{\mathcal{Z}}+\sum_{k>0}a_{k}\cos(\omega_{k}t),\label{eq:logLE}
\end{equation}
 where $a_{k}=(1-c_{k}^{-1})\sin^{2}(\Delta\theta_{k})/2$ which at
this order is equivalent to $a_{k}=(1-c_{k}^{-1})(\Delta\theta_{k})^{2}/2$
and $\omega_{k}=2\Lambda_{k}^{1}$. Similarly, for the logarithm of
the normalized and linearized quantity, $\mathcal{Z}_{\mathrm{F}}(t) = \ln \left[ d_{\mathrm{eff}}\mathcal{L}(t) \right]$, we have 
\begin{equation}
\mathcal{Z}_{\mathrm{F}}(t)=\overline{\mathcal{Z}_{\mathrm{F}}}+\sum_{k>0}a_{k}^{\mathrm{F}}\cos(\omega_{k}t),\label{eq:logLEf}
\end{equation}
 where $a_{k}^{\mathrm{F}}=(1-c_{k}^{-2})(\Delta\theta_{k})^{2}/2$.
From the very similar form of the coefficients $a_{k}$ and $a_{k}^{\mathrm{F}}$,
we can see that both quantities will have qualitatively the same properties.
If we now assume that the frequencies $\left\{ \omega_{k}\right\} _{k>0}$
are rationally independent, the ergodic theorem implies that both
$\mathcal{Z}\left(t\right)$ and $\mathcal{Z}_{\mathrm{F}}\left(t\right)$
are given by a sum of independent random variables. The situation
is perfectly analogous to that analyzed in Sec.~\ref{sec:Statistics-for-generic}
(cfr.~Eq.~(\ref{eq:Zt})) with the difference that Eqns.~(\ref{eq:logLE})
and (\ref{eq:logLEf}) represent a sum of $L/2$ independent variables
as opposed to $D-1$ for Eq.~(\ref{eq:Zt}). This is clearly due
to the quasi-free character of the model.

As in the general case (cfr.~Eq.~(\ref{eq:char_general})),
rational independence allows to compute exactly the characteristic
function of the centered variable $\mathcal{Z}\left(t\right)-\overline{\mathcal{Z}}$:\begin{equation}
\overline{e^{i\lambda\left(\mathcal{Z}\left(t\right)-\overline{\mathcal{Z}}\right)}}=\prod_{k>0}J_{0}\left(\left|\lambda a_{k}\right|\right).\end{equation}

All considerations given in Sec.~\ref{sec:Statistics-for-generic}
carry over also in this case; however, knowledge of the precise analytical
form of the weights $a_{k}$, $a_{k}^{\mathrm{F}}$ allows for a more
detailed analysis. 

The first result we recall here is that in the limit $L\to\infty$
the central limit theorem holds independently from other parameters
such as quench amplitudes, temperature, and so on. In other words,
when $L$ is the largest length scale of the system (the off-critical
region) the variable $\left(\mathcal{Z}-\overline{\mathcal{Z}}\right)/\sqrt{L}$
tends in distribution to a Gaussian (the same result holds of course
also for the linearized version). To show this, just note that since
$0\le a_{k},a_{k}^{F}\le1/2$, the total variance grows extensively
in the whole parameter region. In fact, using $J_{0}\left(x\right)=1-x^{2}/4+O\left(x^{4}\right),$
one obtains that the variance of $\mathcal{Z}$ is $\kappa_{2}=\left(1/2\right)\sum_{k>0}a_{k}^{2}$.
Now, for $L$ large one has: $\kappa_{2}\simeq L/\left(4\pi\right)\int_{0}^{\pi}\left(a_{k}\right)^{2}\le L/8$
meaning that the variance always grows extensively with $L$. This
in turns implies that, for $L\to\infty$, $\left(\mathcal{Z}-\overline{\mathcal{Z}}\right)/\sqrt{L}$
is Gaussian-distributed with mean zero and variance $\sigma^{2}=1/\left(4\pi\right)\int_{0}^{\pi}\left(a_{k}\right)^{2}$.
For the sake of the reader, we also compute the variance of $\mathcal{L}$
in Appendix \ref{sec:Computation-of-variances}.

\begin{figure}
\begin{centering}
\includegraphics[width=7cm]{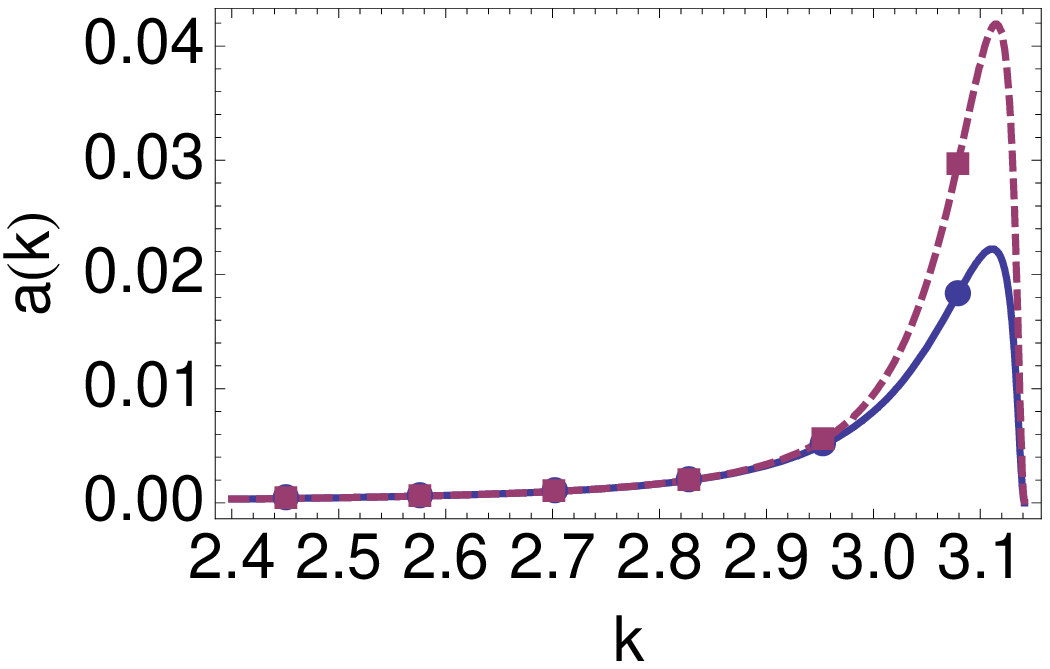} 
\par\end{centering}

\begin{centering}
\includegraphics[width=7cm]{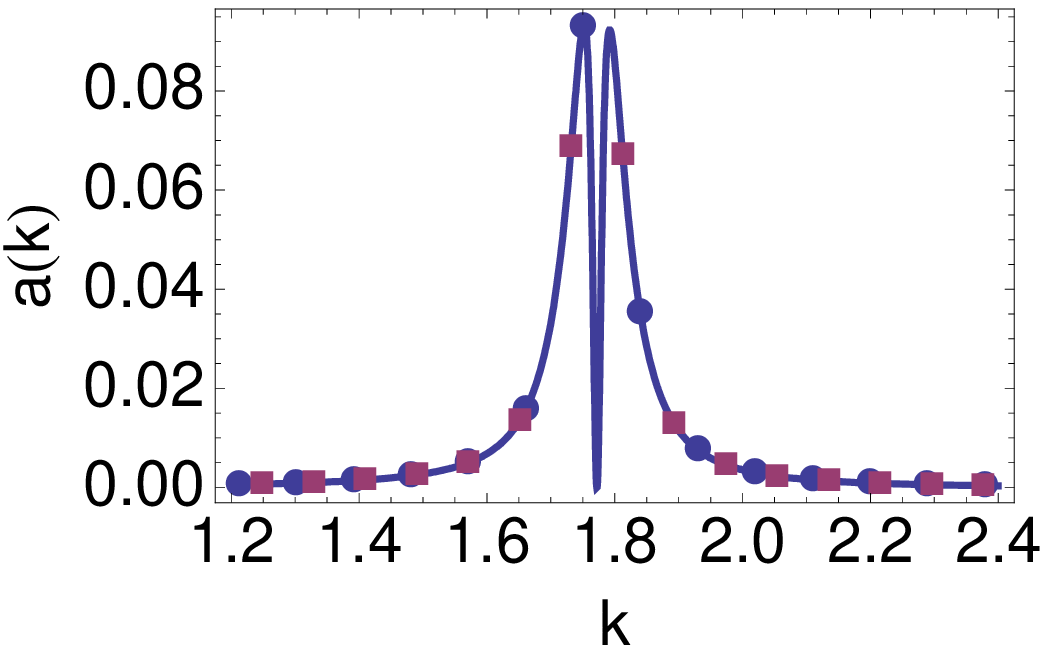} 
\par\end{centering}

\caption{Top panel: $a_{k}$ (solid) and $a_{k}^{F}$ (dashed), for parameters $\gamma^{0,1}=1$,
$h^{0}=0.99,\, h^{1}=1.01$, $\beta=16.67$ ($T=0.06$). The dots
and squares give the allowed quasi-momenta for $L=50$. Bottom panel:
$a_{k}$ for near the anisotropy transition, with parameters $h^{0,1}=0.2$,
$\gamma^{0}=0.01,\,\gamma^{1}=-0.01$, and $\beta=40$ ($T=0.025$).
Circles show allowed weights for $L=70$ while squares refer to $L=78$.
In the latter case two weights $a_{j}$ have approximately the same
value. This results in a single-peaked distribution function.\label{fig:ak}}

\end{figure}

On the other hand, let us now keep $L$ finite and concentrate on
the zero-temperature quasi-critical region, i.e.~$T=0,\,\xi\gg L$.
As for general models, the weights $a_{k}$ become highly peaked in
the quasi-critical region, with very few dominating terms, so that
Eqns.~(\ref{eq:logLE}) and (\ref{eq:logLEf}) correspond to a sum
of \emph{few} random variables, thus invalidating the conditions needed
for the CLT to hold. This is clearly visible in Fig.~\ref{fig:ak},
where we plot the coefficients $a_{k}$ for small quenches close to
the Ising and anisotropic critical lines. From Fig.~\ref{fig:ak},
we can see the conditions in which few momenta contribute: sufficiently
widely-spaced quasi-momenta, i.e.~small-enough system size as compared
to the width of the peak. As we will show explicitly, this corresponds
to the quasi-critical region. 

Let us now turn on the temperature. In the XY model considered here,
$\zeta=\nu=1$ for both kinds of transitions. According to the discussion
in Sec. \ref{sub:Temperature-and-criticality}, the divergence of
$a_{k}$ at low energy is immediately suppressed for temperatures
larger than the gap. Correspondingly, we expect a double-peaked distribution
for $\mathcal{Z}$ whenever the condition $T\ll\Delta$ is satisfied
(the gap $\Delta$ is given in this case by the smallest $\omega_{k}$).
These general findings are confirmed by the explicit analysis of the
temperature-dependent weights $a_{k}\left(T\right)$. Specifically
$a_{k}\left(T\right)=d_{\omega_{k}}\left(T\right)a_{k}\left(T=0\right),$
where the temperature-damping factor $d_{\omega}\left(T\right)=1-\cosh\left(\omega/T\right)^{-m}$
($m=1,2$ for the LE and linearized LE respectively). Since $d_{\omega_{k}}\left(T\right)\approx1$
for $\Delta/T\gg1$, the $T=0$ behavior is recovered when $T\ll\Delta$.
The expansion of $a_{k}\left(T\right)$ in the region $\Delta/T\ll1$
confirms that the $T=0$ divergence is suppressed. On the contrary,
for large temperatures the effect of $d_{\omega}\left(T\right)$ is
that of damping low-energy levels with respect to high-energy ones,
resulting in more evenly-distributed weights $a_{k}$ and making more
pronounced the Gaussian behavior of $\mathcal{Z}$.

To summarize, the small-quench scenario is the following: i) quasi-critical,
low temperature region $\xi\gg L,\, T\ll\Delta$ $\Rightarrow$ few
dominating weights $a_{k}$ $\Rightarrow$ largely spread, double-peaked
distribution for $\mathcal{Z}$; ii) off-critical region, $L\gg\xi$
$\Rightarrow$ large number of dominating weights $a_{k}$ $\Rightarrow$
CLT and Gaussian behavior for $\mathcal{Z}$. An intermediate regime
corresponds to an interpolation between these two limiting distributions
(Gaussian and double-peaked).

\begin{figure}
\begin{centering}
\includegraphics[width=8.4cm]{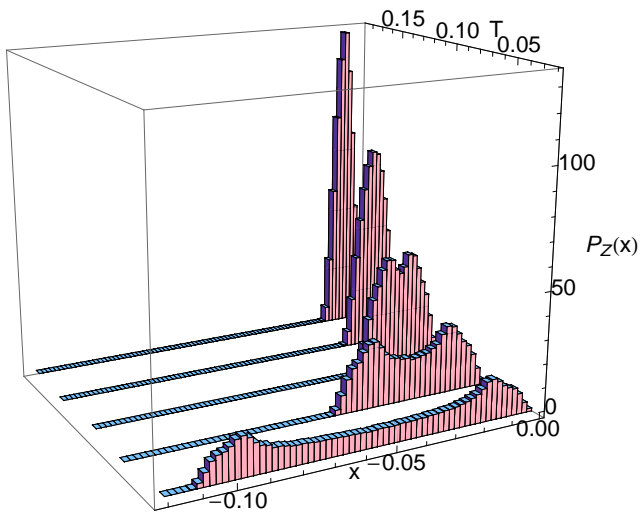} 
\par\end{centering}

\caption{$P_{\mathcal{Z}}(x)$ as the temperature is varied for a quench near
the Ising transition. Notice how the double-peaked distribution becomes
more Gaussian with an increase of temperature. Here, the temperature
takes values $T=0.02,0.06,0.1,0.14,0.18$ and the other parameters
are $L=50$, $h^{0}=0.99$, $h^{1}=1.01$, and $\gamma^{0,1}=1$.
\label{fig:TempDependence}}

\end{figure}

The factor $\sin^{2}(\Delta\theta_{k})$ may be written more-explicitly
in terms of the Hamiltonian parameters as \begin{equation}
\sin^{2}(\Delta\theta_{k})=\frac{\sin^{2}(k)}{(\Lambda_{k}^{0}\Lambda_{k}^{1})^{2}}\left[(\gamma^{1}-\gamma^{0})\cos(k)+(\gamma^{1}h^{0}-\gamma^{0}h^{1})\right]^{2}\label{eq:SM_sin2_factor}\end{equation}
 Note that from this expression we can immediately observe that the
Loschmidt echo will be unity for any initial and final quench field
$h^{0}$, $h^{1}$ of the isotropic ($\gamma=0$) model. This follows
from the fact that $\left[H_{0},H_{1}\right]=0$, provided $\gamma^{0}=0=\gamma^{1}$.
In the following we proceed to a detailed analysis of the $\sin^{2}(\Delta\theta_{k})$
factor for small quenches close to both kinds of transition of the
model, in order to characterize the number of terms contributing to
the oscillatory part of Eqns.~(\ref{eq:logLE}) and (\ref{eq:logLEf}).
This analysis closely mirrors that of Ref.~\cite{lcv2010a}.

\subsection{Ising transition}

Consider quenches near the Ising transition, taking $\gamma=1$ with
the difference $\delta h=h^{1}-h^{0}$ assumed to be small. We first
perform a change into energy variables, using Eq.~(\ref{eq:SM_sin2_factor}),
\begin{eqnarray}
\sin^{2}(\Delta\theta_{k}) & = & \frac{\sin^{2}(k)(h^{1}-h^{0})^{2}}{(\Lambda_{k}^{0})^{2}(\Lambda_{k}^{1})^{2}}\\
 & \approx & \frac{(\omega^{2}-E_{m}^{2})(E_{M}^{2}-\omega^{2})\delta h^{2}}{4(h^{0})^{2}\omega^{4}}\\
 & = & c(\omega)\end{eqnarray}
 where $\omega=\Lambda_{\kappa}^{(0)}$, $E_{m}=|1-h^{0}|$, and $E_{M}=|1+h^{0}|$.

As also shown in \cite{lcv2010a}, $c(\omega)$ is a bell-shaped function.
The number of momenta that are included in the peak determines how
many oscillatory terms contribute to $\mathcal{Z}(t)$, and hence
the shape of the distribution $P_{\mathcal{Z}}(x)$. Notice, importantly,
that the shape of the multiplicative factor $(1-c_{k}^{-m})$ ($m=1,2$
for LE and linearized LE, respectively) serves \emph{only} to weight
less-strongly the lower-energy modes. In other words, it appears that
increasing temperature will never concentrate the spectral weights,
but will rather tend to make the distribution more approximately Gaussian
due to the enhanced relative weights of the higher-energy modes.

The width of the peak $\delta c$ of $c\left(\omega\right)$, estimated
by the inflection point, is approximately $\delta c\approx1.8\left|1-h^{0}\right|$
\footnote{Note the different convention for $\Lambda_{k}$ used here, as one-half
that of Ref.~\cite{lcv2010a}.%
}\cite{lcv2010a}. In order for the temperature-dependent pre-factor
$d_{\omega}(T)$ to not smear out the weights towards large frequencies,
the region for which $d_{\omega}\left(T\right)$ is small must be
smaller than $\delta c$. That is, provided the quench parameters
are those that would produce a double-peaked distribution in the $T=0$
case, in order to get a double-peaked distribution at finite temperature
we need $\vert1-h^{0}\vert\beta\gg1$, that is $T\ll\Delta$ ($\Delta$, the gap, is $\vert1-h^{0}\vert$ for $\gamma=1$).

To check this analysis, let's take a look at the following example.
Let $L=50$, $h^{0}=0.99$, $h^{1}=1.01$ and $\gamma^{0}=\gamma^{1}=1$.
In this case, $L\ll|h-1|^{-1}$, so we expect to see a double-peaked
shape in the $T=0$ limit. This is certainly the case, and the distribution
has been plotted in Fig.~\ref{fig:TempDependence}. As the temperature
is increased, the double peak begins to become less evident between
$T=0.1$ and $T=0.14$. As $T$ increases through this range, the
condition $1/\beta\gg\vert h-1\vert$ begins to hold and $P_{\mathcal{Z}}(x)$
becomes more nearly Gaussian.

\subsection{Anisotropy transition}

\begin{figure}
\begin{centering}
\includegraphics[width=8.4cm]{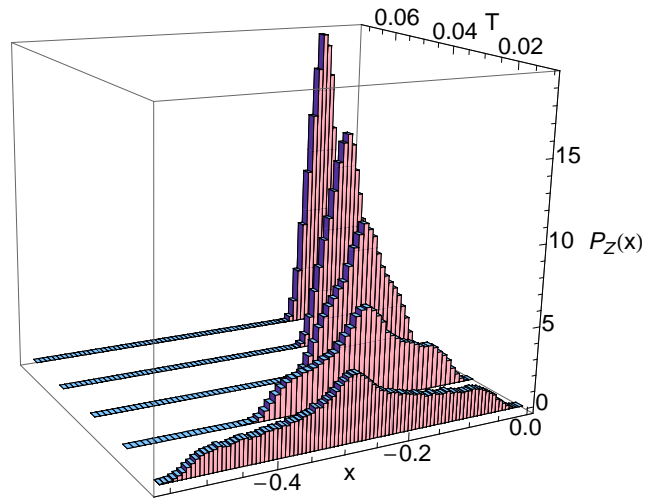} 
\par\end{centering}

\caption{$P_{\mathcal{Z}}(x)$ as the temperature is varied for a quench near
the anisotropy transition. Notice how the single-peaked distribution
becomes more Gaussian with an increase of temperature. Here, the temperature
takes values $T=0.01,0.025,0.04,0.055,0.07$ and the other parameters
are $L=78$, $h^{0,1}=0.2$, $\gamma^{0}=0.01$, and $\gamma^{1}=-0.01$.
\label{fig:TempDependenceAni}}

\end{figure}

The procedure for examining $P_{\mathcal{Z}}(x)$ in the vicinity
of the anisotropy transition goes much the same as for the Ising transition.
The only extra complication here is due to the form of the single-particle
energy, as it is not one-to-one with the momentum and does not
allow an immediate change from momentum to energy variables over the
whole range of possible momenta. Consider the simple case $h^{1}=h^{0}$,
such that we quench only via the anisotropy parameter. In the following
we consider the case $h=0$, where the minimum of the single-particle
energy is obtained for $k_{F}=\pi/2$ and the gap is simply given
by $\Delta=|\gamma|$. For other values of the magnetic field $h$,
the location of the Fermi momentum will shift but the following analysis
will be qualitatively similar. Though $k\to\Lambda_{k}$ is not 1-1,
let us restrict ourselves to the interval $k\in\left[\pi/2,\pi\right]$,
in which the map is indeed 1-1. In this case, define

\begin{equation}
c_{a}(\omega_{0},\omega_{1})=\frac{(1-\omega_{0}^{2})(\omega_{0}^{2}-(\gamma^{0})^{2})}{(1-(\gamma^{0})^{2})(\omega_{0}\omega_{1})^{2}}(\gamma^{1}-\gamma^{0})^{2}\end{equation}
 as the function corresponding to $c(\omega)$ for the Ising model
case. Taking $\vert\delta\gamma\vert\ll1$ and expanding in $\delta\gamma$
to second-order, we obtain \begin{equation}
c_{a}(\omega)=\frac{(1-\omega^{2})(\omega^{2}-(\gamma^{0})^{2})}{(1-(\gamma^{0})^{2})\omega^{4}}(\delta\gamma)^{2}.\end{equation}
 Just as in the Ising case, this function is bell-shaped. The width
of $c_{a}\left(\omega\right)$ as given by the location of the inflection
point is approximately $\delta c_{a}\approx1.8\left|\gamma^{0}\right|$.
Exactly as found near the Ising transition, the temperature factor
suppresses the peak of the $\sin^{2}(\Delta\theta)$, broadening the
number of momenta which contribute appreciably to the sums (\ref{eq:logLE})
and (\ref{eq:logLEf}). Consequently, in addition to the $T=0$ requirements
for obtaining a double-peaked distribution, we must require that $1/\beta\ll|\gamma^{0}|$, 
i.e. $T\ll\Delta$. One notable difference between the statistics
for quenches near the anisotropy versus Ising transition is the potential
for obtaining a single-peaked distribution such as is shown in Fig.
\ref{fig:TempDependenceAni}, directly analogous to the single-particle
DOS of a two-dimensional isotropic tight-binding model. This phenomenon
is due to the band structure near the anisotropy transition in which
the Fermi momentum is at an intermediate (incommensurate) value of
$k\in\left[0,\pi\right]$ rather than at the edge, as it is near the Ising
transition. Consequently, the weights $a_{k}$ have a double-peaked
form, as shown in the bottom panel of Fig.~\ref{fig:ak}. For certain
system sizes, then, it is possible that the quasi-momenta will be
approximately symmetrically placed about the Fermi momentum, resulting
in the weights $a_{k}$ appearing in pairs (see e.g. the $L=78$ example
in the bottom panel of Fig.~\ref{fig:ak}). In the quasi-critical
regime, since the separation between peaks of the double-peaked distribution
for $P_{\mathcal{Z}}(x)$ is proportional to the difference between
the two largest weights (peaks of $P_{\mathcal{Z}}(x)$ are at $\overline{\mathcal{Z}}\pm\vert a_{1}-a_{2}\vert$,
where $a_{1,2}$ are the two largest weights \cite{lcv2010a}), if
the two maximal weights are nearly equal the peaks will appear merged.
However, this distribution is to be understood as simply a special
case of the usual double-peaked behavior we have observed for the
Ising case. Indeed, if we were to take the same set of parameters
for the above example but a slightly different system size (e.g. $L=70$),
we will obtain the same type of double-peaked distribution as we have
seen in Fig. \ref{fig:TempDependence}. We remark that the $h^{0,1}=0$
case we have considered in the preceding analysis is slightly pathological,
in that due to the Fermi momentum appearing at exactly $k_{F}=\pi/2$,
all weights $a_{k}$ will come in pairs. However, for non-zero magnetic
fields we find that in the overwhelming majority of cases we will
find a double-peaked distribution as the quasi-momenta corresponding
to the maximal pair of weights will normally not be very symmetrically-displaced
about the Fermi level.

\section{Conclusion}

In this paper we have studied the problem of the long-time out-of-equilibrium dynamics of a quantum system after a Hamiltonian quench
of a thermal (Gibbs) state. More specifically, we have examined the
infinite-time statistics of the thermal Loschmidt echo $\mathcal{L}$
\cite{CamposVenuti2011}. This latter is a global quantity that depends
just on the Hamiltonian and on the initial state, i.e. no preferred
observable has to be singled out, and bears relevance for the general
quantum equilibration problem. $\mathcal{L}$ can be naturally defined
as the Uhlmann fidelity between the initial finite-temperature Gibbs
state associated with a Hamiltonian and the time-dependent one obtained
by the unitary evolution corresponding to a different (quenched) Hamiltonian.

For a small quench we argued, on fairly general grounds, that a Log-Normal
distribution $P_{\mathcal{L}}$ is realized for off-critical systems
at arbitrary temperatures. On the other hand, if the quantum quench
is performed near a quantum critical point a dramatically different
scenario emerges. For sufficiently small temperatures a universal
double-peaked distribution isomorphic to the DOS of a two-dimensional
tight-binding model is realized. As the temperature increases this
singular distribution can either be continuously morphed into the
former Log-Normal distribution or not depending on whether or not the
quench enacts a sufficiently irrelevant perturbation.

In the second part of the paper we applied the above general analysis
to the paradigmatic case of a finite-temperature quantum XY chain
in a transverse magnetic field. In this case a plethora of analytical
results can be obtained for arbitrarily large quenches, far or close
to the critical lines of the model. In particular, we have found a
tight lower bound on the Loschmidt echo for a class of XY-type models
in terms of a simplified {}``linearized'' Loschmidt echo.

The extension of this long-time probability distribution approach
to general physically relevant observables appears as a compelling
task for future investigations. 

\emph{Acknowledgments-} We acknowledge useful discussions with
Siddhartha Santra. NTJ is grateful for support from an Oakley Fellowship,
LCV acknowledges support from European project COQUIT under FET-Open
grant number 2333747, and PZ acknowledges support from NSF grants
PHY-803304, PHY-0969969, and DMR-0804914.

\bibliographystyle{apsrev} \bibliographystyle{apsrev} \bibliographystyle{apsrev}
\bibliographystyle{apsrev} \bibliographystyle{apsrev}
\bibliography{ref_le}

\appendix

\section{Proof of the qubit inequality\label{sec:bound--qubit}}

We prove here the following inequality for arbitrary single-qubit
states $\rho$ and unitary operators $U$, \begin{equation}
\frac{\mathrm{Tr}\left[U\rho U^{\dagger}\rho\right]}{\mathrm{Tr}\left[\rho^{2}\right]}\leq\mathrm{Tr}\left[\sqrt{\rho^{\frac{1}{2}}U\rho U^{\dagger}\rho^{\frac{1}{2}}}\right]^{2}.\label{appendix_qubit_inequality}\end{equation}
 Expressing $\rho$ in Bloch form, $\rho=\frac{1}{2}\big(\1+\vec{v}\cdot\vec{\sigma}\big)$,
we have \begin{eqnarray*}
\mathrm{Tr}\left[\rho_{0}\rho_{1}\right] & = & \frac{1}{2}\big(1+\vec{v_{0}}\cdot\vec{v_{1}}\big),\\
\mathrm{det}\left[\rho\right] & = & \frac{1}{4}\big(1-v^{2}\big).\end{eqnarray*}
 As shown by H\"ubner \cite{Hubner1992}, the Uhlmann fidelity between
arbitrary single-qubit states $\rho$ and $\sigma$ is \[
\mathrm{Tr}\left[\sqrt{\rho^{\frac{1}{2}}\sigma\rho^{\frac{1}{2}}}\right]^{2}=\mathrm{Tr}\left[\rho\sigma\right]+2\sqrt{\mathrm{det}\left[\rho\right]\mathrm{det}\left[\sigma\right]}.\]
 Using this and the fact that a unitary transformation leaves the
magnitude of the Bloch vector unchanged, define $v=|v_{0,1}|$, $\vec{v_{0}}\cdot\vec{v_{1}}=v^{2}\cos{\theta}$,
and \begin{eqnarray*}
\mathrm{RHS} & = & \mathrm{Tr}\left[\rho^{2}\right]\mathrm{Tr}\left[\sqrt{\rho^{\frac{1}{2}}U\rho U^{\dagger}\rho^{\frac{1}{2}}}\right]^{2}\\
 & = & \frac{1}{2}(1+v^{2})\left[\frac{1}{2}(1+v^{2}\cos\theta)+\frac{1}{2}(1-v^{2})\right],\\
\mathrm{LHS} & = & \mathrm{Tr}\left[U\rho U^{\dagger}\rho\right]\\
 & = & \frac{1}{2}(1+v^{2}\cos\theta).\end{eqnarray*}
 Finally, \begin{eqnarray*}
4(\mathrm{RHS}-\mathrm{LHS}) & = & v^{2}(1-\cos\theta)-v^{4}(1-\cos\theta)\\
 & = & v^{2}(1-\cos\theta)(1-v^{2})\\
 & \geq & 0,\end{eqnarray*}
 since $0\leq v^{2}\leq1$.

\section{Proof of the bound for quasi-free fermions\label{sec:bound-XY}}

We now prove that the inequality (\ref{eq:qubit_inequality}) holds
for Gibbs states of quasi-free fermions of the form Eq. (\ref{eq:Quasi-Free}):
\begin{equation}
\frac{\mathrm{Tr}(U\rho U^{\dagger}\rho)}{\mathrm{Tr}(\rho^{2})}\leq\left[\mathrm{Tr}\sqrt{\rho^{\frac{1}{2}}U\rho U^{\dagger}\rho^{\frac{1}{2}}}\right]^{2}\label{eq:SM_Lower_bound}\end{equation}
Exploiting the tensor-product form for the Gibbs state \[
\rho^{\alpha}=\frac{e^{-\beta H_{\alpha}}}{Z^{\alpha}}=\bigotimes_{k>0}\frac{1}{Z_{k}^{\alpha}}\left[\rho_{k}^{\alpha}\oplus\1_{k}\right],\]
 where \cite{PZ07} \begin{eqnarray*}
J_{k}^{\alpha} & := & \cos(\theta_{k}^{\alpha})\sigma_{k}^{z}+\sin(\theta_{k}^{\alpha})\sigma_{k}^{y}\\
\rho_{k}^{\alpha} & := & \exp(-\beta\Lambda_{k}^{\alpha}J_{k}^{\alpha})=\cosh(\beta\Lambda_{k}^{\alpha})\1-\sinh(\beta\Lambda_{k}^{\alpha})J_{k}^{\alpha}\end{eqnarray*}
 we can straightforwardly compute \begin{eqnarray*}
\mathrm{Tr}\left[U\rho U^{\dagger}\rho\right] & = & \prod_{k>0}\frac{1}{Z_{k}^{2}}\left[\mathrm{Tr}\left[U_{k}\rho_{k}U_{k}^{\dagger}\rho_{k}\right]+2\right],\\
\mathrm{Tr}\left[\rho^{2}\right] & = & \prod_{k>0}\frac{1}{Z_{k}^{2}}\left[\mathrm{Tr}\left[\rho_{k}^{2}\right]+2\right],\\
\mathrm{Tr}\left[\sqrt{\rho^{\frac{1}{2}}U\rho U^{\dagger}\rho^{\frac{1}{2}}}\right]^{2} & = & \prod_{k>0}\frac{1}{Z_{k}^{2}}\left[\mathrm{Tr}\left[\sqrt{\rho_{k}^{\frac{1}{2}}U_{k}\rho_{k}U_{k}^{\dagger}\rho_{k}^{\frac{1}{2}}}\right]+2\right]^{2}.\end{eqnarray*}
 Note that $Z_{k}=2(1+\cosh(\beta\Lambda_{k}))$, so $4\leq Z_{k}<\infty$.
For the following, define $x:=\beta\Lambda_{k}^{0}$ and $v:=\sin^{2}(\beta\Lambda_{k}^{1}t)(1-\cos(2\Delta\theta_{k}))$,
where $0\leq v\leq2$. We also make use of the following: \begin{eqnarray*}
\mathrm{Tr}\left[U_{k}\rho_{k}U_{k}^{\dagger}\rho_{k}\right] & = & 2\cosh(2x)-2\sinh^{2}(x)v,\\
\mathrm{Tr}\left[\rho_{k}^{2}\right] & = & 2\cosh(2x),\\
\mathrm{Tr}\left[\sqrt{\rho_{k}^{\frac{1}{2}}U_{k}\rho_{k}U_{k}^{\dagger}\rho_{k}^{\frac{1}{2}}}\right]^{2} & = & 2\left[2\cosh^{2}(x)-\sinh^{2}(x)v\right]\end{eqnarray*}
 Defining $f(x,v):=2\cosh^{2}(x)-\sinh^{2}(x)v$, we have \begin{eqnarray*}
\mathrm{Tr}\left[U_{k}\rho_{k}U_{k}^{\dagger}\rho_{k}\right] & = & 2(f(x,v)-1),\\
\mathrm{Tr}\left[\sqrt{\rho_{k}^{\frac{1}{2}}U_{k}\rho_{k}U_{k}^{\dagger}\rho_{k}^{\frac{1}{2}}}\right]^{2} & = & 2f(x,v).\end{eqnarray*}
 Now, taking only the $k^{\mathrm{th}}$ term of the products above,
in order to prove the validity of (\ref{eq:SM_Lower_bound}) we need
to check whether the quantity \begin{eqnarray*}
Q(x,v) & := & \frac{1}{Z_{k}^{2}}\left[\mathrm{Tr}\left[\rho_{k}^{2}\right]+2\right]\left[\mathrm{Tr}\left[\sqrt{\rho_{k}^{\frac{1}{2}}U_{k}\rho_{k}U_{k}^{\dagger}\rho_{k}^{\frac{1}{2}}}\right]+2\right]^{2}\\
 & - & \left[\mathrm{Tr}\left[U_{k}\rho_{k}U_{k}^{\dagger}\rho_{k}\right]+2\right]\end{eqnarray*}
 is non-negative. Re-writing in terms of $f(x,v)$, as above, we have
\begin{eqnarray*}
Q(x,v) & = & \frac{\cosh(2x)+1}{2(1+\cosh(x))^{2}}\Big(\sqrt{2f(x,v)}+2\Big)^{2}-2f(x,v).\end{eqnarray*}

Consider \begin{eqnarray*}
\partial_{v}Q(x,v) & = & \frac{-\cosh(2x)+1}{(1+\cosh(x))^{2}}\sinh^{2}(x)\left[1+\frac{\sqrt{2}}{2\sqrt{f(x,v)}}\right]\\
 &  & +2\sinh^{2}(x),\end{eqnarray*}
 and \begin{eqnarray*}
\partial_{v}^{2}Q(x,v) & = & \frac{-\sqrt{2}(\cosh(2x)+1)\sinh^{4}(x)}{2(1+\cosh(x))^{2}}\Big(f(x,v)\Big)^{-3/2}\\
 & \leq & 0.\end{eqnarray*}
 Since the concavity of $Q(x,v)$ with respect to $v$ is never positive,
to minimize it we need only consider the boundaries, $v=0,2$. We
find that $Q(x,0)=0$ and $Q(x,2)\geq0$ $\forall x$, and hence have
shown that $Q(x,v)\geq0$ $\forall(x,v)$. The inequality (\ref{eq:SM_Lower_bound})
follows.

\section{Effective dimension $d_{\mathrm{eff}}$\label{sec:Effective-dimension}}

The effective dimension is defined to be the reciprocal of the purity,
$d_{\mathrm{eff}}=\mathrm{Tr}\left[\rho^{2}\right]^{-1}$. Now, the
purity of the Gibbs state (\ref{Gibbs_state}) is simply \begin{eqnarray*}
\mathrm{Tr}\left[\rho^{2}\right] & = & \mathrm{Tr}\left[\bigotimes_{k>0}\frac{1}{Z_{k}^{2}}\left[\rho_{k}^{2}\oplus\1_{k}\right]\right]\\
 & = & \prod_{k>0}\frac{1}{Z_{k}^{2}}\Big(\mathrm{Tr}\left[\rho_{k}^{2}\right]+2\Big)\\
 & = & \prod_{k>0}\frac{2+2\cosh(2\beta\Lambda_{k})}{4(1+\cosh(\beta\Lambda_{k}))^{2}}\\
 & = & \prod_{k>0}\Big(\frac{\cosh(\beta\Lambda_{k})}{1+\cosh(\beta\Lambda_{k})}\Big)^{2}.\end{eqnarray*}

\section{Linearized Loschmidt echo $\mathcal{L}_{\mathrm{F}}(t)$\label{sec:Linearized-Loschmidt-echo}}

The {}``linearized'' Loschmidt echo is defined to be \begin{eqnarray*}
\mathcal{L}_{\mathrm{F}}(t) & = & \mathrm{Tr}\left[\rho(t)\rho\right]\\
 & = & \mathrm{Tr}\left[\bigotimes_{k>0}\frac{1}{(Z_{k}^{0})^{2}}\left[e^{-itH_{k}^{1}}e^{-\beta H_{k}^{0}}e^{itH_{k}^{1}}e^{-\beta H_{k}^{0}}\oplus\1_{k}\right]\right]\\
 & = & \prod_{k>0}\frac{1}{(Z_{k}^{0})^{2}}\left[2+\mathrm{Tr}\left[e^{-itH_{k}^{1}}e^{-\beta H_{k}^{0}}e^{itH_{k}^{1}}e^{-\beta H_{k}^{0}}\right]\right],\end{eqnarray*}
 where $H_{k}^{0,1}$ are the Hamiltonian operators on the even subspace
of the $k^{\mathrm{th}}$ momentum subsystem (See \cite{PZ07}). Now,
to evaluate the trace inside the product, we need the following \cite{PZ07}:
\begin{eqnarray*}
e^{-itH_{k}^{1}} & = & \cos(\Lambda_{k}^{1}t)-i\sin(\Lambda_{k}^{1}t)J_{k}^{1}\\
e^{-\beta H_{k}^{0}} & = & \cosh(\beta\Lambda_{k}^{0})-\sinh(\beta\Lambda_{k}^{0})J_{k}^{0},\end{eqnarray*}
 and \begin{eqnarray*}
\mathrm{Tr}\left[J_{k}^{0}J_{k}^{1}\right] & = & 2\cos(\theta_{k}^{0}-\theta_{k}^{1})\\
\mathrm{Tr}\left[(J_{k}^{1}J_{k}^{0})^{2}\right] & = & 2\cos(2(\theta_{k}^{0}-\theta_{k}^{1}))\end{eqnarray*}
 After some algebra, we have \begin{eqnarray*}
\mathcal{L}_{\mathrm{F}}(t) & = & \prod_{k>0}\left[\Big(\frac{\cosh(\beta\Lambda_{k}^{0})}{1+\cosh(\beta\Lambda_{k}^{0})}\Big)^{2}\right.\\
 & - & \left.\Big(\frac{\cosh(\beta\Lambda_{k}^{0})-1}{\cosh(\beta\Lambda_{k}^{0})+1}\Big)\sin^{2}(\Delta\theta_{k})\sin^{2}(\Lambda_{k}^{1}t)\right]\\
 & = & \mathrm{Tr}\left[\rho^{2}\right]\prod_{k>0}\left[1-\Big(1-\frac{1}{\cosh^{2}(\beta\Lambda_{k}^{0})}\Big)\times\right.\\
 &  & \left.\sin^{2}(\Delta\theta_{k})\sin^{2}(\Lambda_{k}^{1}t)\right]\end{eqnarray*}
 The time average $\overline{\mathcal{L}_{\mathrm{F}}}$ can be straightforwardly
calculated by exploiting the rational independence of the single-particle
energies.

\section{Time-average, $\overline{\mathcal{L}}$\label{sec:Computation-of-averages}}

To compute the time average, re-group the product in the expression
(\ref{eq:XY_Loschmidt_echo}) for the Loschmidt echo into a sum: \[
\mathcal{L}(t)=1+\sum_{k>0}X_{k}(t)+\sum_{k_{1}>k_{2}>0}X_{k_{1}}(t)X_{k_{2}}(t)+...\]
 where \begin{eqnarray*}
X_{k}(t) & = & \sum_{m=1}^{\infty}h_{k}^{(m)}\sin^{2m}(\Lambda_{k}^{1}t)\end{eqnarray*}
 and \[
h_{k}^{(m)}:=\left\{ \begin{array}{ll}
\frac{c_{k}b_{k}}{1+c_{k}}, & m=1\\
\frac{2c_{k}}{(1+c_{k})^{2}}(b_{k})^{m}\binom{1/2}{m}, & m>1\end{array}\right.\]
 Note that \[
\overline{\sin^{2m}(x)}=2^{-2m}\binom{2m}{m}=(-1)^{m}\binom{-1/2}{m}.\]
 Making use of the rational independence of the single-particle energies
$\lbrace\Lambda_{k}\rbrace_{k>0}$ and regrouping into a product,
we obtain \[
\overline{\mathcal{L}}=\prod_{k>0}\left(1+G_{k}^{(1)}\right),\]
 with \begin{eqnarray*}
G_{k}^{(1)} & = & \sum_{m=1}^{\infty}h_{k}^{(m)}(-1)^{m}\binom{-1/2}{m}\\
 & = & \frac{c_{k}b_{k}}{2(1+c_{k})}+\frac{2c_{k}}{(1+c_{k})^{2}}\left[\frac{2}{\pi}\textrm{E}(-b_{k})-\frac{b_{k}}{4}-1\right],\end{eqnarray*}
 where \begin{eqnarray*}
c_{k} & := & \cosh(\beta\Lambda_{k}^{0})\\
b_{k} & := & -\big(1-c_{k}^{-2}\big)\sin^{2}(\Delta\theta_{k})\end{eqnarray*}
 and $\textrm{E}(x)$ is the complete elliptic integral.

\section{Variance of $\mathcal{L}$\label{sec:Computation-of-variances}}

For completeness we sketch here the procedure to compute the variances.
Squaring Eq. (\ref{eq:XY_Loschmidt_echo}), regrouping the product
into a sum, and taking the time average in exactly the same way as
we did for the mean, we obtain 

\begin{equation}
\overline{\mathcal{L}^{2}}=\prod_{k>0}\left(1+G_{k}^{(2)}\right),\end{equation}
 where \begin{equation}
G_{k}^{(2)}=\sum_{m=1}^{\infty}g_{k}^{(m)}(-1)^{m}\binom{-1/2}{m}\end{equation}
 and \begin{equation}
g_{k}^{(m)}=2h_{k}^{(m)}+\sum_{n=1}^{m-1}h_{k}^{(n)}h_{k}^{(m-n)}.\end{equation}
 We now consider small quenches, expanding $\Delta^{2}\mathcal{L}:=\overline{\mathcal{L}^{2}}-(\overline{\mathcal{L}})^{2}$
to lowest order in $\Delta\theta_{k}$ with \emph{fixed} system size
$L$. (The determination of whether the quench is {}``large'' or
{}``small'' must be made with respect to the system size). Expanding
the square mean and the mean-squared terms into sums, \begin{eqnarray*}
\overline{\mathcal{L}^{2}} & = & 1+\sum_{k>0}G_{k}^{(2)}+\sum_{k_{1}>k_{2}>0}G_{k_{1}}^{(2)}G_{k_{2}}^{(2)}+\dots\\
(\overline{\mathcal{L}})^{2} & = & 1+\sum_{k>0}\Big(2G_{k}^{(1)}+(G_{k}^{(1)})^{2}\Big)\\
 &  & +4\sum_{k_{1}>k_{2}>0}G_{k_{1}}^{(1)}G_{k_{2}}^{(1)}+\dots\end{eqnarray*}
 Only the lowest non-zero power of $\Delta\theta_{k}$ is to be retained,
which turns out to be the fourth-order term. We find, \begin{eqnarray*}
\Delta^{2}\mathcal{L} & \approx & \sum_{k>0}\Big(G_{k}^{(2)}-2G_{k}^{(1)}-(G_{k}^{(1)})^{2}\Big)\\
 &  & +\sum_{k_{1}>k_{2}>0}G_{k_{1}}^{(2)}G_{k_{2}}^{(2)}-4G_{k_{1}}^{(1)}G_{k_{2}}^{(1)}\\
 & \approx & \frac{1}{8}\sum_{k>0}\Big(1-\frac{1}{\cosh(\beta\Lambda_{k}^{0})}\Big)^{2}(\Delta\theta_{k})^{4}.\end{eqnarray*}
 We can see from this expression how increasing the temperature (lowering
$\beta$) tends to decrease the variance, while if the angle differences
$\Delta\theta_{k}$ are large (as occurs for even small quenches near
a critical point), the variance becomes larger. Note that here we
have fixed $L$, so this small-quench expansion does not provide information
on the finite-size scaling of the variance. Since each term in the
products defining $\overline{\mathcal{L}}^{2}$ and $\overline{\mathcal{L}^{2}}$
is smaller than unity, generically each of these products will vanish
exponentially as a function of $L$. Hence, since the variance is
a difference of two exponentially small functions we expect the variance
to vanish exponentially with $L$ as well. 
\end{document}